# Supernova II Neutrino Bursts and Neutrino Massive Mixing*

UCLA-APH-0131-12/00

David B. Cline

*Department of Physics and Astronomy, Box 951547*
*University of California Los Angeles*
*Los Angeles, CA 90095-1547, USA*

**Abstract**. We describe the Neutrino Spectrum and detection for SN II sources. We discuss the effects of neutrino mixing in the SN II. A new analysis of SN1987A is described. We discuss the possible detection of the diffuse relic SN II flux. Finally we discuss a new detection concept, OMNIS, for $\nu_\mu$ and $\nu_\tau$ and detection and compare with other present and future SN detectors.

## 1. INTRODUCTION OF THE NEUTRINO SPECTRUM FROM A SN II

The issue of whether or not neutrinos have masses is important for astrophysics and cosmology. Astrophysical considerations may represent the best hope for determining neutrino masses and mixings. In this paper, we examine how proposed neutral-current-based, supernova neutrino-burst detectors, in conjunction with the next generation water-Čerenkov detectors, could use a galactic supernova event to either measure or place constraints on the $\nu_{\mu,\tau}$ masses in excess of 5 eV[1,2]. Such measurements would have important implications for our understanding of particle physics, cosmology, and the solar neutrino problem and would be complementary to proposed laboratory vacuum-oscillation experiments.

**Table 1**

| | Tests | | | | | Nucleosynthesis | |
|---|---|---|---|---|---|---|---|
| **Scheme** | **$\nu_\odot$** | **$\nu_{atmos}$** | **LBL** | **SBL** | **SN ν's** | **BBN** | **SNN** |
| I<br>3ν mixing<br>No LSND | Yes<br>$\nu_e \to \nu_{\mu,\tau}$ | Yes<br>$\nu_\mu \to \nu_\tau$ | Yes<br>$\nu_\mu \to \nu_\tau$ | No<br>$\nu_\tau \to \nu_e$<br>$\nu_\mu \to \nu_e$<br>τ appearance? | √ | OK<br><br>$\nu_\mu \to \nu_e$ | OK |
| II<br>4ν mixing<br>$\nu_\mu \to \nu_\tau$<br>(Doublet)<br>$\nu_\odot \to \nu_s$<br>LSND | Yes<br>$\nu_e \to \nu_s$<br>(No extra<br>N.C. signal) | Yes<br>$\nu_\mu \to \nu_\tau$ | Yes<br>$\nu_\mu \to \nu_\tau$ | No?<br>$\nu_e \to \nu_\tau$<br>τ appearance?<br>$\nu_e \to \nu_\tau$?<br>$\nu_e \to \nu_\mu$? | √<br>Hot<br>?<br>Maybe!<br>Maybe! | <br>D/N<br>$\nu_e$-spectrum<br>? | ??<br>Good or bad<br>?? |
| III<br>4ν mixing<br>$\nu_\mu \to \nu_s$<br>Doublet<br>No LSND | Yes<br>$\nu_e \to \nu_\mu$ | Yes<br>$\nu_\mu \to \nu_s$ | Yes<br>$\nu_\mu \to \nu_s$ | No<br>$m_{\nu\tau}$<br>$\nu_\mu \to \nu_\tau$? | √<br><br><br>? | ?<br><br>T of F | r-process<br>constraint |

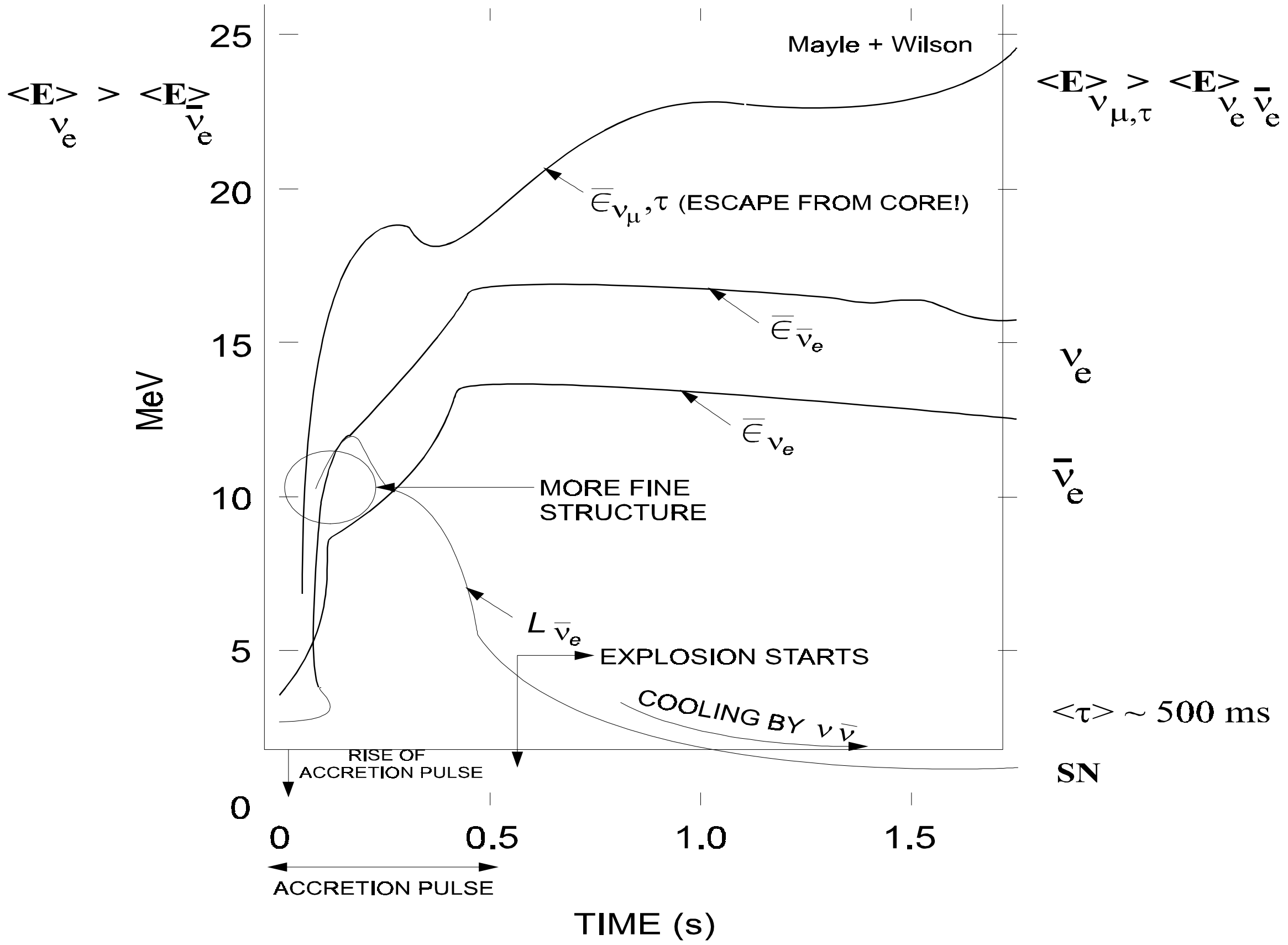

Fig. 1. The expected neutrino luminosity and average energy of the different species following the work of Mayle and Wilson.

A light neutrino mass between 1 eV and 100 eV would be highly significant for cosmology. In fact, if a neutrino contributes a fraction $\Omega_\nu$ of the closure density of the Universe, it must have a mass $m_\nu \approx 92\ \Omega_\nu\ h^2$ eV, where $h$ is the Hubble parameter in units of 100 km $s^{-1}$ $Mpc^{-1}$. Reasonable ranges for $\Omega_\nu$ and $h$ then give 1 eV to 30 eV as a cosmologically significant range. A neutrino with a mass in the higher end of this range (*i.e.*, $10 \leq m_\nu \leq 30$ eV) could contribute significantly to the closure density of the Universe. The cosmic background explorer (COBE) observation of anistropy in the microwave background, combined with observations at smaller scales, and the distribution of galaxy streaming velocities, have been interpreted as implying that there are two components of dark matter: hot ($\Omega_{HDM} \sim 0.3$) and cold ($\Omega_{CDM} \sim 0.6$). The hot dark matter (HDM) component could be provided by a neutrino with a mass of about 7 eV.[1]

## 2. EFFECTS OF NEUTRINO MIXING FOR SN II

SN II emit all types of neutrinos and anti-neutrinos with about the same luminosity per neutrino on anti-neutrino species. The expected energy spectrum is very different as can be seen in Fig. 2. Table 1 gives a set of options

[1]While current cosmological date do not favor this, it is not completely excluded.

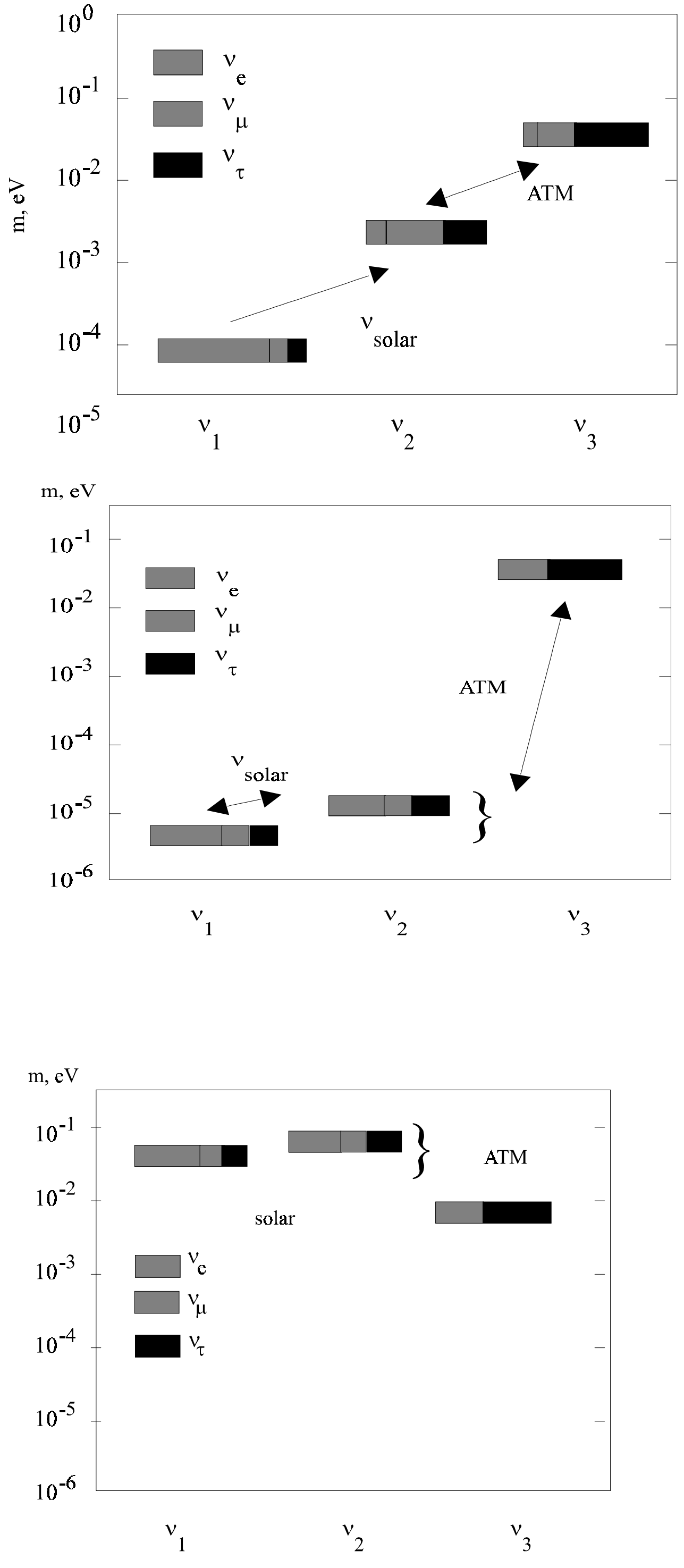

$\nu$

Fig. 2a, 2b, 2c. Different schemes for the Neutrino Mass spectrum adopted from Ref. 3.

for mixing and tests.

The possible different models of neutrino mass is shown in Fig. 2. These different mass structures have important consequences for the MSW effect in the SN II and the search for the correct neutrino mass spectrum.

## 3. A NEW ANALYSIS OF SN1987A AND THE LMA SOLAR NEUTRINO SOLUTION

Because the $\nu_x$ spectrum is expected to be harder than the $\nu_e$ or $\bar{\nu}_e$ (Fig. 3), one signature for neutrino oscillation is to observe a hard component in the $\nu_e$ or $\bar{\nu}_e$ spectrum from the process $\nu_x \to \nu_e$ or $\bar{\nu}_x \to \bar{\nu}_e$ [2]. We can define the neutrino flux in the following way:

$$\langle \nu_e \rangle = (1 - p)\langle \nu_e \rangle_o + p\langle \nu_x \rangle_o \quad , \qquad (1)$$

$$\langle \bar{\nu}_e \rangle = (1 - \bar{p})\langle \bar{\nu}_e \rangle_o + \bar{p}\langle \bar{\nu}_x \rangle_o \quad , \qquad (2)$$

where $\langle \nu_e \rangle_o$, $\langle \bar{\nu}_e \rangle_o$, $\langle \nu_x \rangle_o$, $\langle \bar{\nu}_x \rangle_o$ denote the unmixed neutrino spectra and $p, \bar{p}$ the mixing fraction.

In vacuum, we can write $p = (1/2) \sin^2 2\theta$ and $\bar{p} = (1/2) \sin^2 2\theta$. Figure 3 shows the distorted $\bar{\nu}_e$ event spectrum for various values of $\bar{p}$. Thus, even a small mixing ($\bar{p} = 0.2$) causes an appreciable event spectrum distortion at high energy and should be readily detected in the next supernova event.

As is well known, there were 20 events recorded in SN1987A: 12 by the Kamiokande detector [4] and 8 by the IMB detector [5]. We first turn to the initial analysis of L. Krauss [6].

First we comment on the Kamiokande and IMB event populations:

1. The IMB detector had a strong bias against low energy events.
2. The mass of the IMB detector was about three times larger than the Kamiokande detector and thus was more sensitive to higher energy neutrino events that are less probable.
3. The Kamiokande detector had excellent low energy properties, as was later demonstrated by the observation of solar neutrinos.
4. Some of the pmt's for the IMB detector were not operational during the recording of SN1987A events.

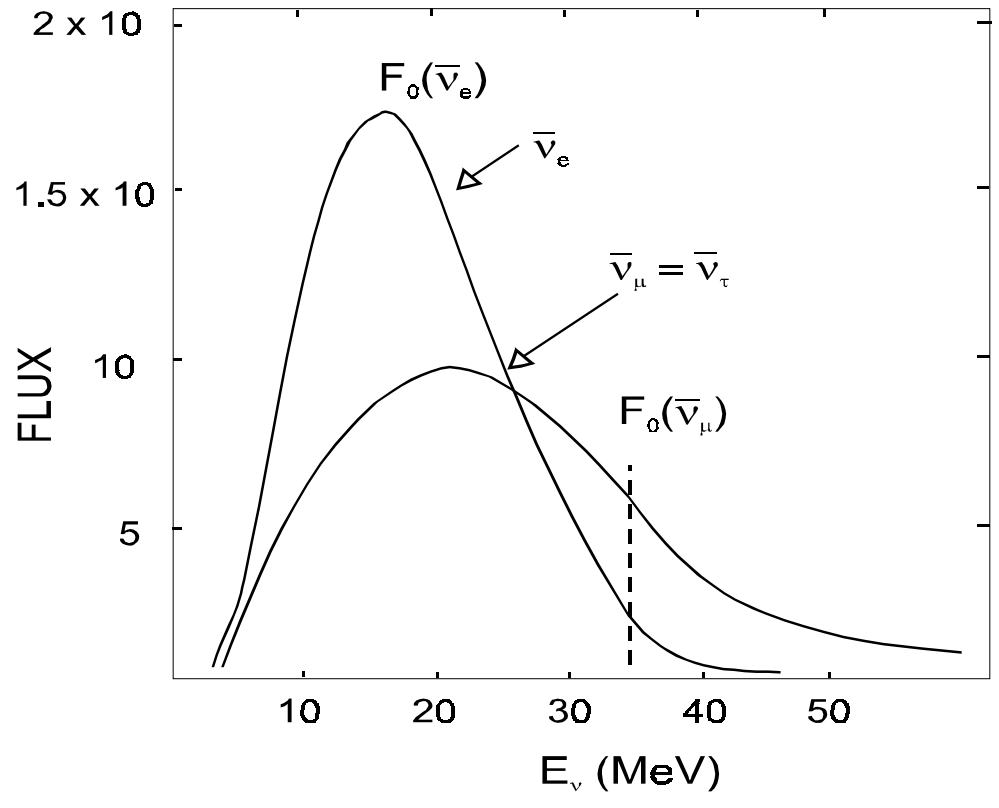

Figure 3. Expected $\bar{\nu}_e$ and $\bar{\nu}_x$ spectra from an SNII explosion.

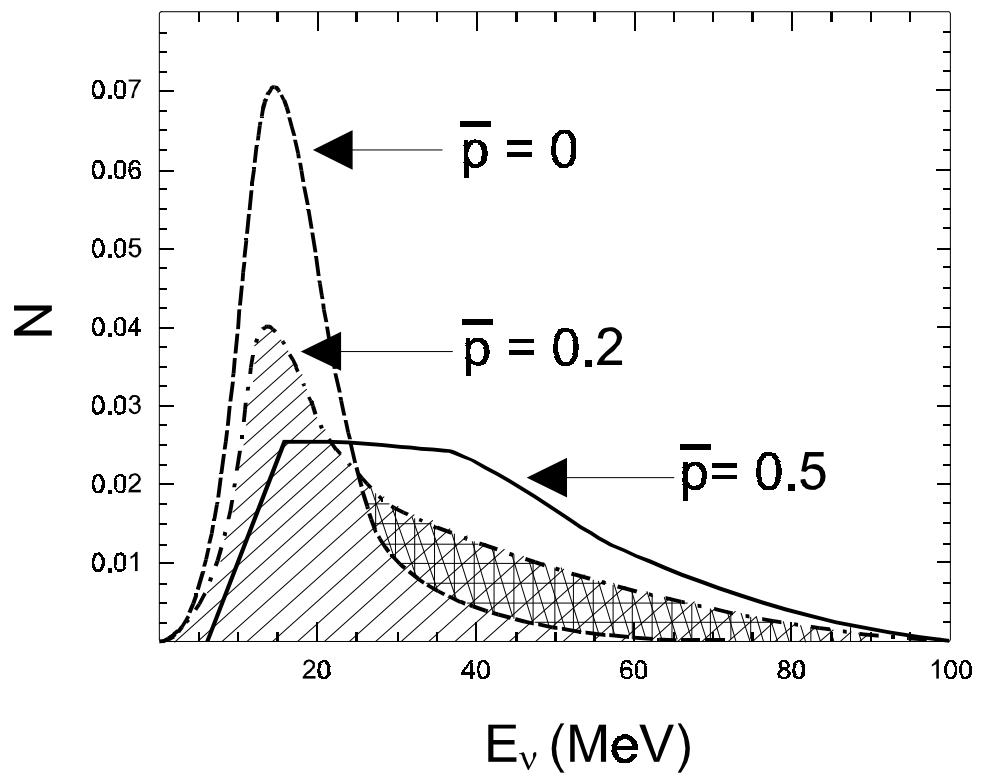

Figure 4. Events distribution expected for different values of the mixing $\bar{p}$.

In the Krauss analysis, an attempt was made to incorporate both populations of events by correcting the model as shown in Fig. 5. While this showed an acceptable fit, the mean temperature of the distribution was 4.5 MeV, which was higher than would be expected normally.

Smirnov, Spergel, and Bahcall tried another analysis [7]. They fit the combined Kamiokande and IMB data with a model that allowed for $\nu_x \to \nu_e$ mixing as given by formula (2). They did not correct for the obvious differences in the Kamiokande and IMB event population (threshold, different detector, masses, possible differencesindetectionefficiency, etc.) but just added up the integral of the total event energies. They gave an exhaustive discussion of the different types of temperature distributions that may occur for the different neutrino flavors from the supernova emission. They expressed their results as a function of the mean energy of the $\nu_x$ neutrinos. Since most models gave this value to be 22 MeV or greater, we use that value in the results shown here. In Fig. 5, we have replotted the results of this analysis for the 95% confidence level reported in the paper. Note that most of the LMA and all of the vacuum oscillation (or "just so") is excluded.

One can be critical of this analysis due to the fact that no attempt was made to correct for the different experimental conditions in the Kamiokande and IMB experiments. However, these results may well be a conservative lower limit, since corrections for the experimental differences will decrease the impact of the high energy events in IMB, as shown by the Krauss analysis.

### 3a. A NEW ANALYSIS OF THE SN1987A DATA FOR NEUTRINO MIXING

Because of the problems of comparing the two populations of events illustrated in points 1 through 4 above, we propose that a sensible analysis should use the data set with the least bias. Based on the Kamiokande data alone, while this set has no event with an energy above 40 MeV, there is no reason why the detector would not have recorded such events had they been produced. In Fig. 6, we show the Kamiokande data and the predictions of neutrino mixing (Fig. 3). The case $\bar{p}$ = 1/2 is excluded to at least 99% confidence level. Even with a lower statistical sample, the conclusions of this analysis are as powerful as those of Smirnov, Spergel, and Bahcall. Table 2 gives the Kolmogurov test for these data [8]. We believe these results largely exclude the LMA solar neutrino solution.

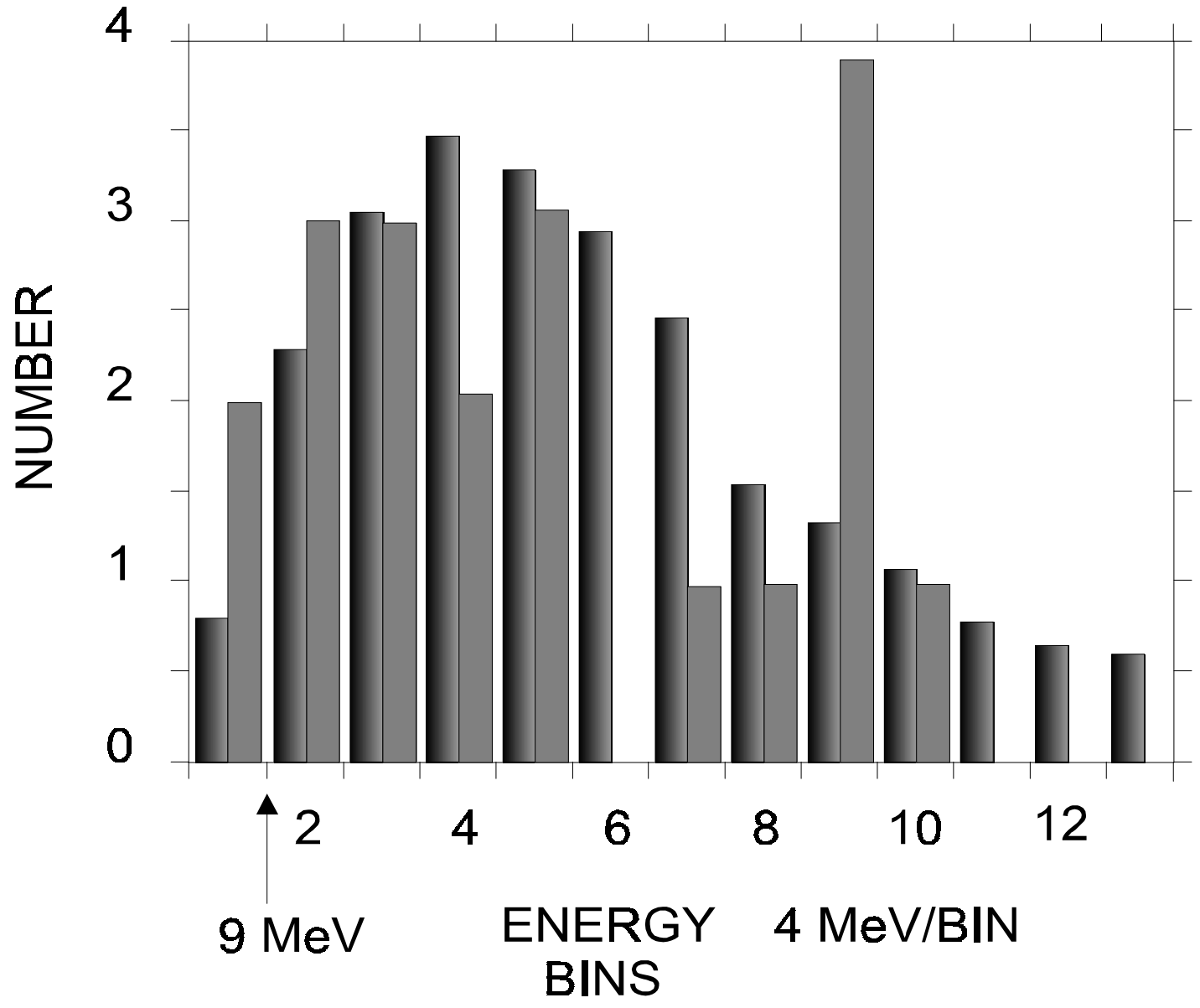

Figure 5. Krauss analysis of SN1987A data. The hatched events are the combined data and the solid blocks are the results of a model that incororates the effects of the detectors, etc. [6]. Neutrino mixing was not assumed.

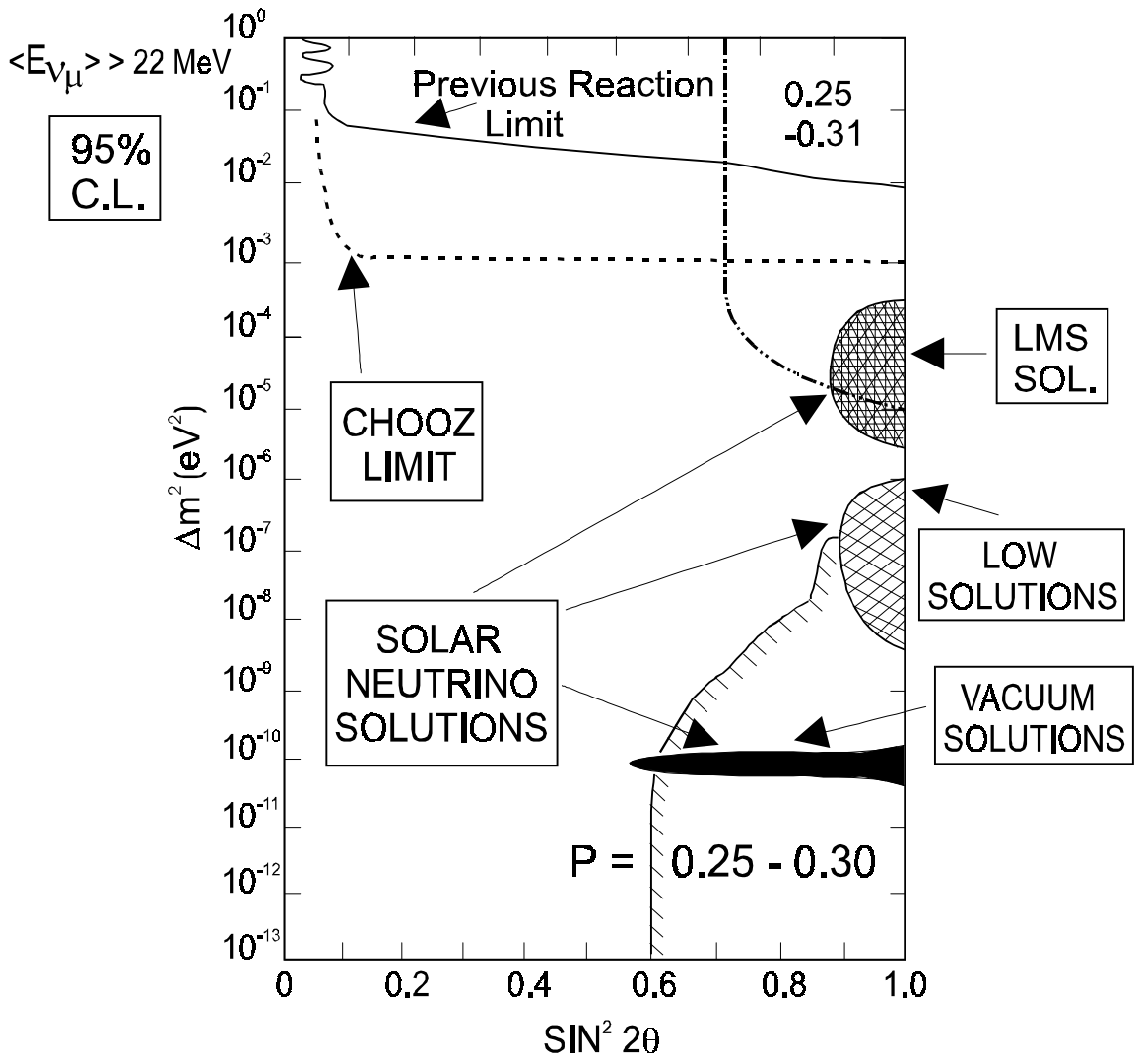

Figure 6. Limits on  from SN1987A (from Ref. [7]).

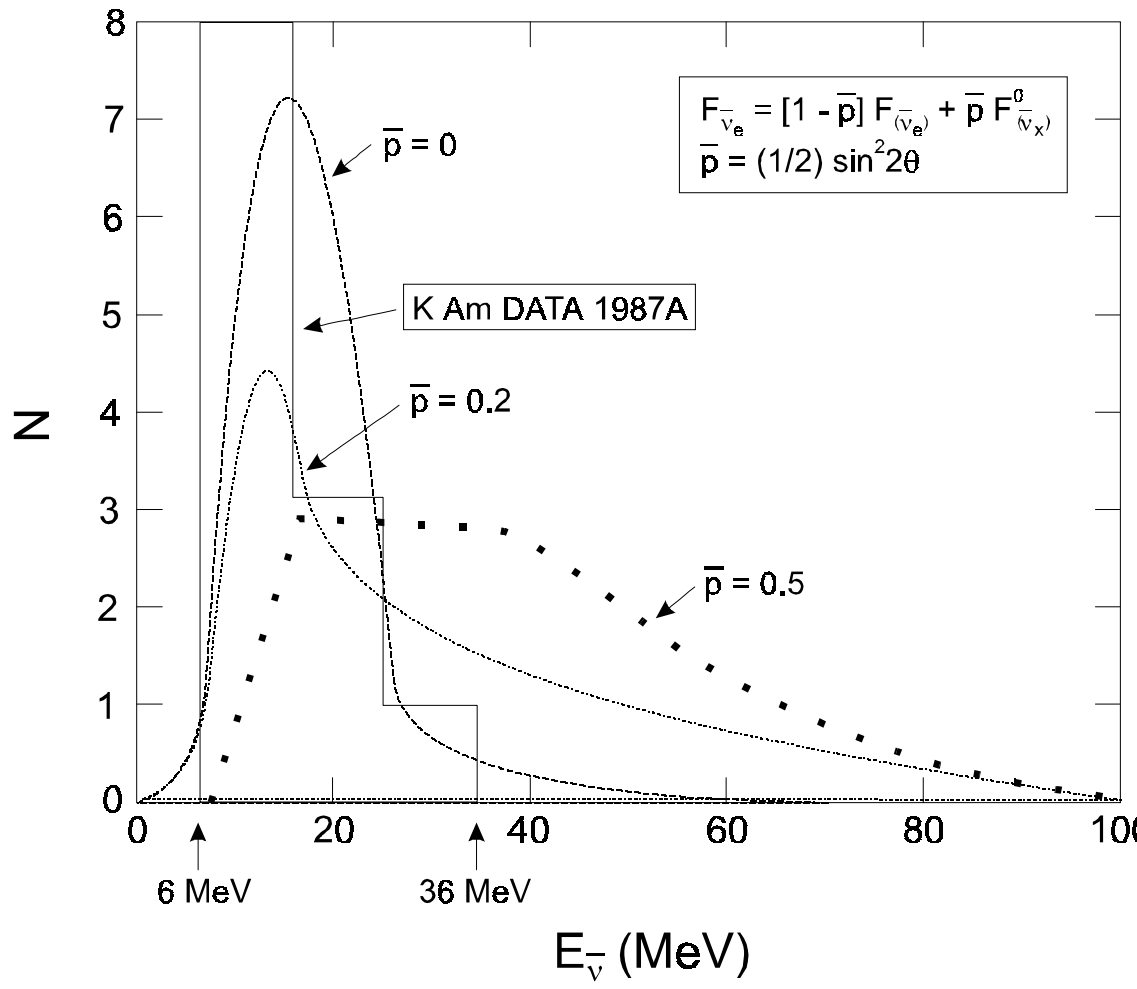

Figure 7. Comparison of the Kamiokande data with the neutrino oscillation models.

**Table 2. Kolmogurov Test for the Model and data in Fig. 8.**

| **Parameter, $\bar{p}$** | **Probability of Hypothesis (%)**<br>**Confidence Level (%)** |
|---|---|
| **0** | **58**<br>**42** |
| **0.2** | **3.6**<br>**96.4 excluded** |
| **0.5** | **0.02**<br>**> 99 excluded** |

## 4. POSSIBLE DETECTION OF THE DIFFERENT RELIC NEUTRINO FLUX FROM PAST SN II

Another kind of relic neutrinos are the neutrinos that arise from the integrated flux from all past type-II supernovae. Figure 8 shows a schematic of these (and other) fluxes. These fluxes could be modified by transmission through the SNII environment, as discussed recently. [1]

The detection of $\bar{\nu}_e$ from the relic supernovae may someday be accomplished by the SK detector. It would be as interesting to detect $\nu_e$ with an ICARUS detector, as illustrated in Table 3. High-energy $\nu_e$ would come from $\nu_{\mu,\tau} \rightarrow \nu_e$ in the supernova. A window of detection occurs between the upper solar neutrino energy and the atmospheric neutrinos, as first proposed by D. Cline and reported in the first ICARUS proposal (1983-1985). The ideal detector to observe this is a large ICARUS liquid-argon detector. [1]

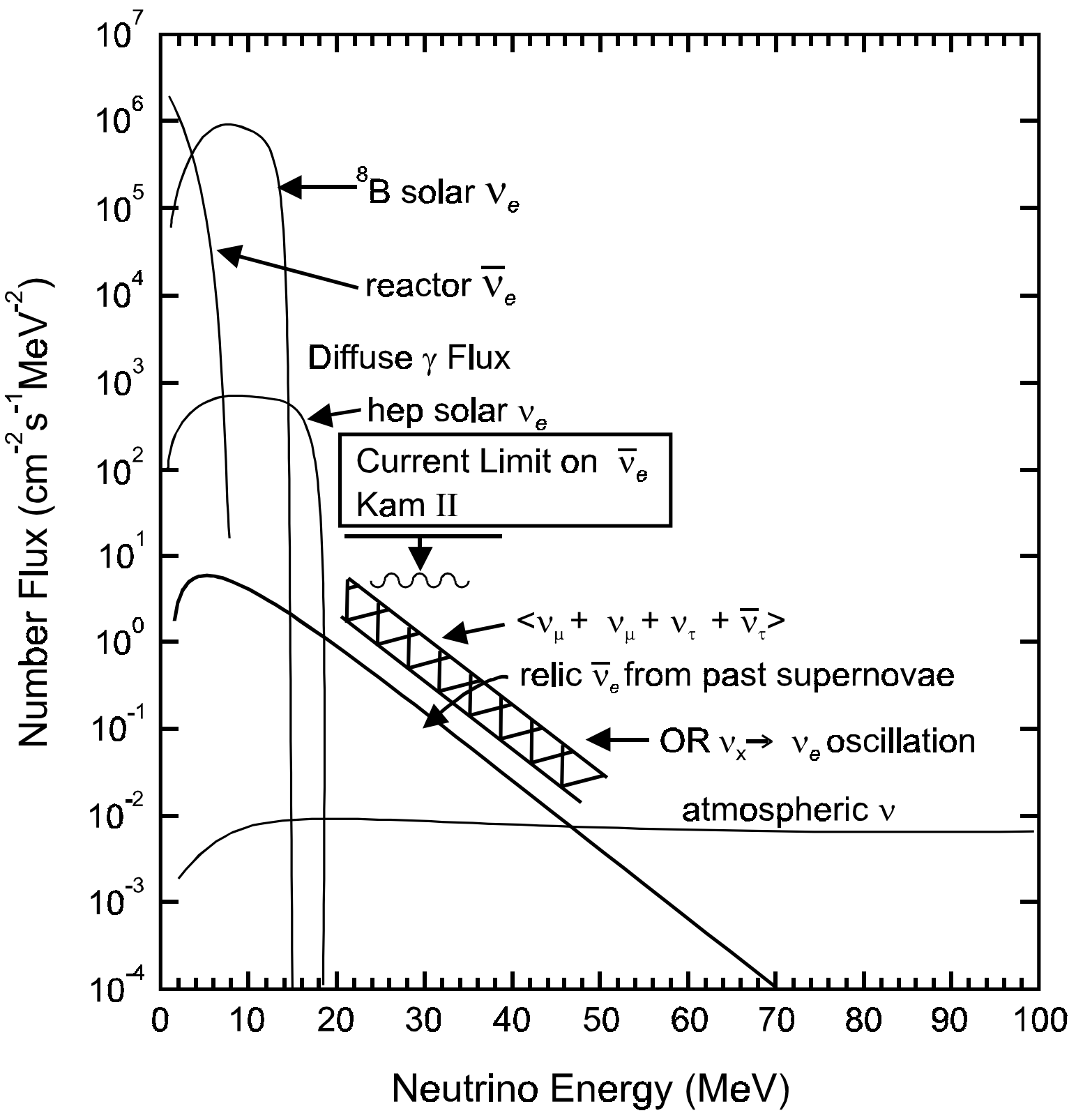

FIGURE 8. Relic neutrinos from past supernova. Note: $\nu_x \rightarrow \nu_e$ in the supernova can boost the energy of the $\nu_e$ if we find $<E\ \nu_e> > <E\bar{\nu}_e>$. This will be a signal for neutrino oscillation in supernovae! and measure $\sin^2 \theta x_e$. [1]

**Table 3. Detection of $\nu/\bar{\nu}_e$ relic neutrino flux from time integrated SNII.**

1. Relic $\nu/\bar{\nu}_e$ from all SNII back to $Z \sim 5$: $<E_\nu> \sim 1/(1 + Z)<E_\nu>$
2. Detection would give integrated SNII rate from Universe
   –Window of detetion [D. Cline, ICARUS proposal, 1984]
3. Neutrino oscillations in SNII would give $\nu_x \rightarrow \nu_e$ with higher energy than $\bar{\nu}_e$
4. Detect $\bar{\nu}_e$ with SK or ICARUS. Attempt to detect $\nu_x/\nu_e$ detection.

## 5. THE OMNIS DETECTOR CONCEPT AND OTHER SUPER NOVA NEUTRINO DETECTORS

Recently there has been real progress in supernova simulations giving an explosion. These calculations give interesting predictions for the neutrino spectra. Detectors like the SK and SNBO/OMNIS may be able to detect such effects, however the SNBO/OMNIS detector may be of crucial importance for this study. Using these various detectors, it should be possible to detect a finite neutrino mass. The characteristics of this detector are listed in Table 4.

In this analysis, we have assumed the existence of a very massive neutral-current detector (the SNBO/OMNIS), which we discuss next [9]. By using these different detectors it will be possible to measure the μ or τ neutrino masses, as shown in Table 3, which could determine a mass to ~ 10 eV [9]. To go to lower mass, we need to use the possible fine structure in the burst; we have shown that it may be possible to reach ~ 3 eV with very large detectors in this case. The detection of two-neutron final stateswould be useful for Pb detection. [10].

**Table 4. Properties of the Pproposed) OMNIS/SNBO Detector**

| | |
|---|---|
| Targets: | NaCℓ (WIPP site)<br>Fe and Pb (Soudan and Boulby sites) |
| Mass of Detectors: | WIPP site ≥ 200 ton<br>Soudan/Boulby sites ≥ 200 ton |
| Types of Detectors: | Gd in liquid scintillator<br>$^6$Li loaded in the plastic scintillators that are read out by scintillating-fiber−PMT system |

**The major problem of supernova detection is the uncertain period of time between such processes in this** Galaxy.

In addition, complimentary detectors should be active when the supernova goes off in order to gain the maximum amount of information possible about the explosion process and neutrino properties.

Lacking an ideal observatory, a group of us have been studying a very large detector, SNBO/OMNIS. Table 4 gives some of the guidelines for this detector. We have located a possible site for the observatory near Carlsbad, NM, which is the WIPP site (shown in Fig. 9). We believe that expectation that the background for a galactic supernova is much smaller than the signal at this site. A schematic of the SNBO/OMNIS detector is shown in Fig. 9.

The basic idea of OMNIS/SNBO is to develop a long lived inexpensive SN II detector that detects $\upsilon_\tau$ and $\upsilon_\mu$ neutrinos by the neutral current process. The Carlsbad site gives a unique long-lasting undersground laboratory.

In Table 5 we give the rates of different detectors for a galactic supernova and Figure 10 shows the rates by neutrino species[1].

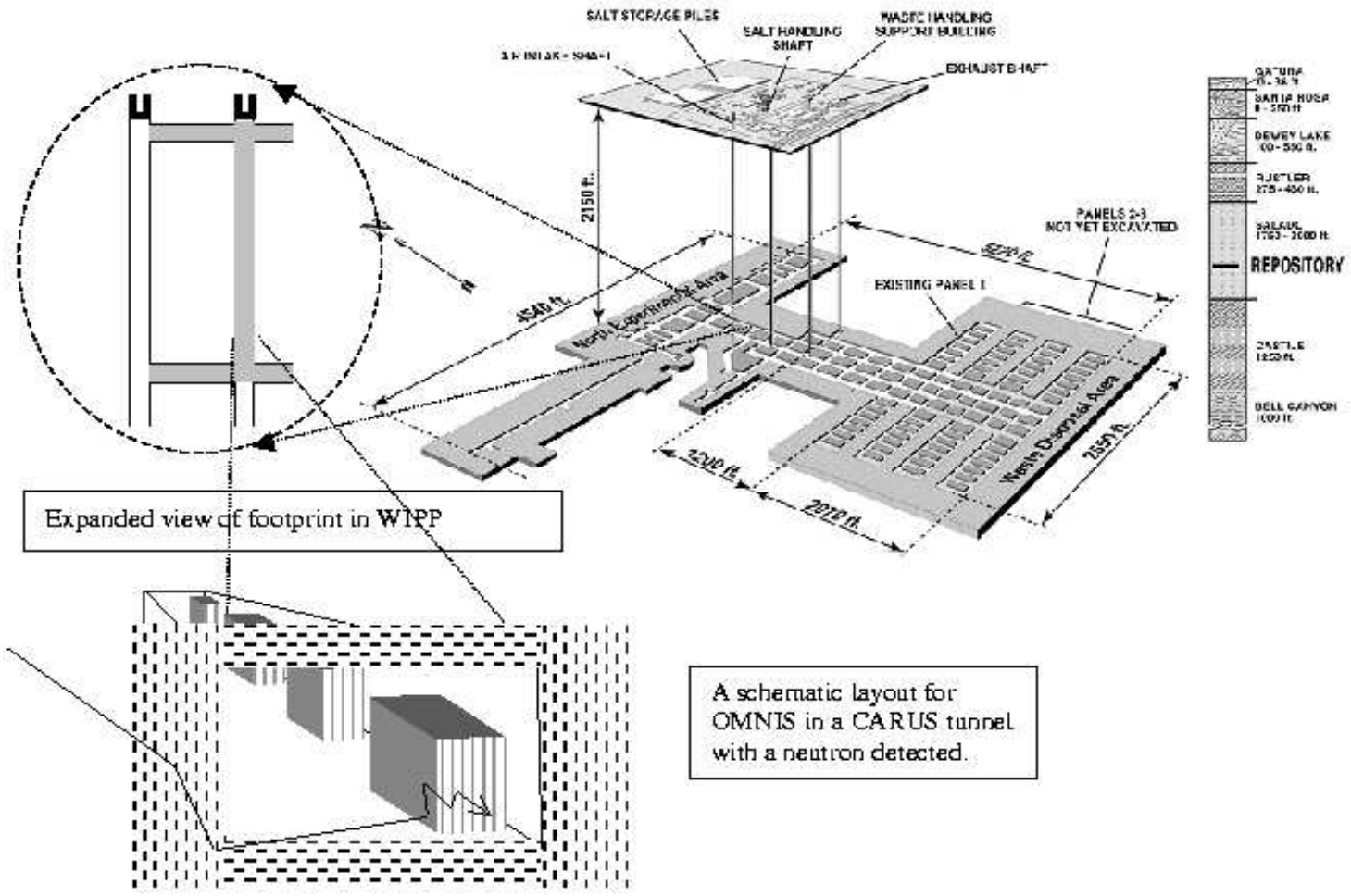

Figure 9. Schematic Layout for OMNIS in a CARUS tunnel with a neutron detected.

**TABLE 5: YIELDS OF SUPERNOVA NEUTRINO DETECTORS**

| Detect\or | Target Material | Fiducial Mass (Ton) | Target Element | Yield $(\upsilon_e)$ | Yield $(\overline{\upsilon_e})$ | Yield $(\upsilon_\mu+\upsilon_\tau+\overline{\upsilon_\mu}+\overline{\upsilon_\tau})$ |
|---|---|---|---|---|---|---|
| Super K | $H_2O$ | 32000 | p, e, O | 180 | 8300 | 50 |
| LVD | $CH_2$ | 1200 | p, e, C | 14 | 540 | 30 |
| SNO | $H_2O$ | 1600 | p, e, O | 16 | 520 | 6 |
| SNO | $D_2O$ | 1000 | d, e, O | 190 | 180 | 300 |
| OMNIS | *Fe* | 8000 | *Fe* | 20* | 20* | 1200* |
| OMNIS | *Pb* | 2000 | *Pb* | | | |
| no osc. | | | | 110** | 40** | 860** |
| $\upsilon_\mu \rightarrow \upsilon_e$ osc. | | | | $\leq 4420$** | 40** | $\leq$ 640** |
| | | | | | | |

* Assumes same efficiency as in Smith 1997
** Assumes a single neutron detection efficiency of 0.6

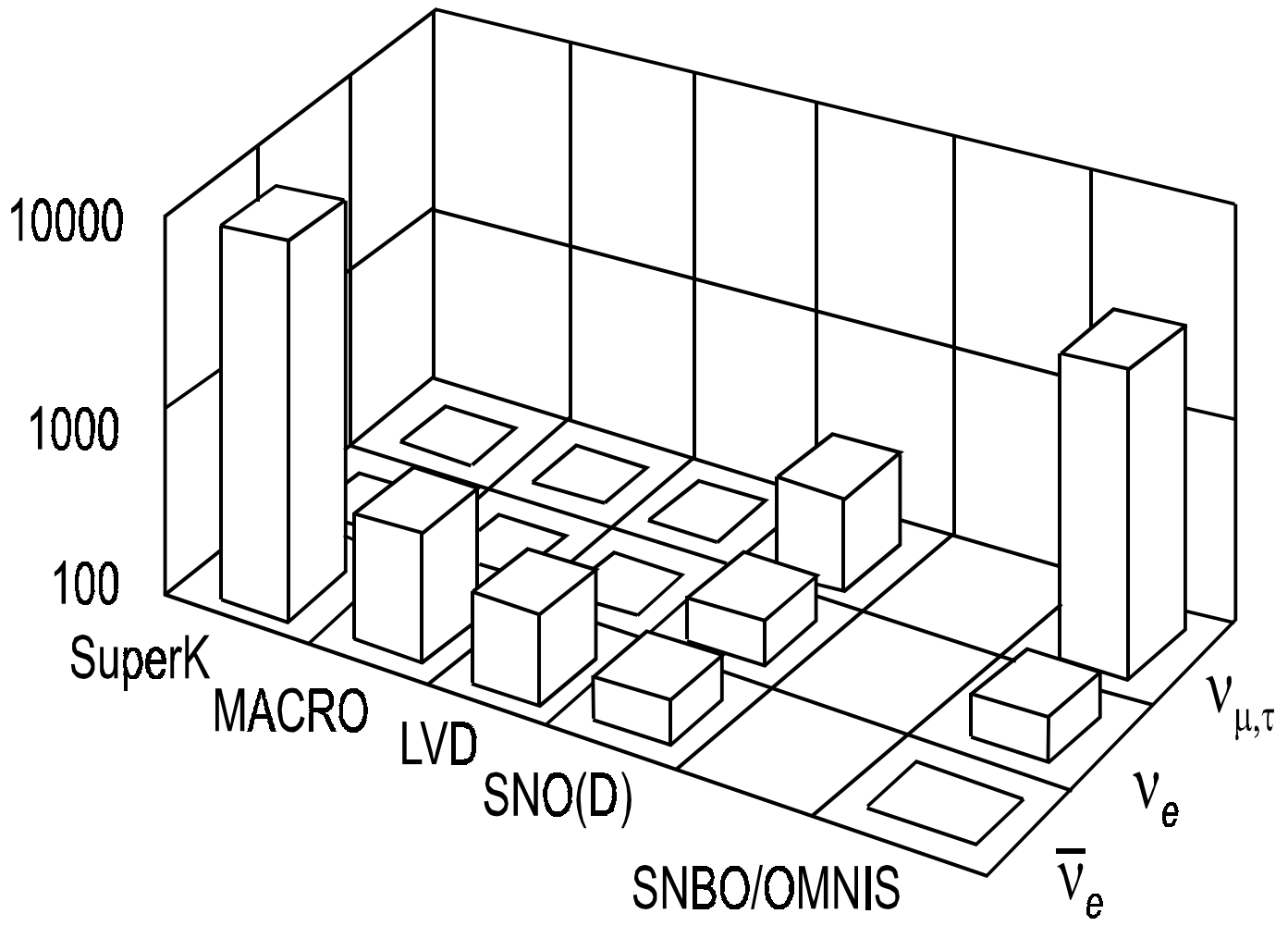

FIGURE 10. Comparison of world detectors (event numbers for supernovae at 8 kpc).

## REFERENCES

1. Cline, D. B. “Search for Relic Neutrinos and Supernova Bursts”, in *Proceedings, Eighth Intl. Wksp. on Neutrino Telescopes* (Venice, Feb. 23-26, 1999), ed. M. Baldo Ceolin, 1999, Vol. II, pp. 309-320.

2. R. Mayle, J. R. Wilson, and D. B. Schramm, Astrophys. J. 318 (1987) 288.

3. Figure 3 is adapted from the paper A. Dighe and A. Yu Smirnov, Phys. Rev. D. **62**, 033007.

4. Kamiokande Collaboration, K. S. Hirata et al., Phys. Rev. Lett. 58 (1987) 1490.

5. IMB Collaboration, R. M. Bionata et al., ibid. 1494.

6. M. Krauss, Nature (London) 329 (1987) 689.

7. A. Yu. Smirnov, D. N. Spergel, and J. N. Bahcall, Phys. Rev. D

8. S. Otwinowski, UCLA, private communication (2000).

9. Cline, D. B., Fuller, G., Hong, W. P., Meyer, B., and J. Wilson, Phys. Rev. D50, 720 (1994).

10. G. M. Fuller, W. C. Haxton, and G. C. McLaughlin, Phys. Rev. D 59 (1999) 085005 (and Refs. therein)